\documentclass[11pt]{article}

\usepackage[T1]{fontenc}
\usepackage[utf8]{inputenc}
\usepackage{microtype}
\usepackage[margin=1in]{geometry}
\usepackage{graphicx}
\usepackage{booktabs}
\usepackage{amsmath}
\usepackage{amssymb}
\usepackage{textcomp}
\usepackage[numbers,sort&compress]{natbib}
\usepackage[colorlinks=true,linkcolor=blue,citecolor=blue,urlcolor=blue]{hyperref}
\usepackage[capitalise]{cleveref}
\usepackage{setspace}
\graphicspath{{figures/}}

\title{\textbf{Large reasoning models for abnormal situation
management in safety-critical industrial processes}}
\author{Khalid Alhazmi\\[2pt]
\small SABIC Technology Center, Al Jubail, 31952, Saudi Arabia\\
\small e-mail: hazmimm@sabic.com}
\date{}

\begin{document}
\maketitle

\begin{abstract}
\noindent
Automation operates safety-critical processes inside their design envelope and
leaves abnormal situations to human operators. Mismanagement of these
situations is a leading contributor to process-safety incidents and a
hindrance to achieving autonomy. Here we show that a general-purpose large
reasoning model, with no task-specific training and only the information
available to an operator, manages abnormal situations at run time through a
bounded, programmatically verified action interface. Across 39 abnormal
situations and operating-point changes on a plant-wide industrial benchmark
process, the reasoning model maintained the plant within all hard constraints
in all 39, while basic regulatory control failed in 15. It matched the plant's
expert-engineered advanced control and diagnosed the root-cause fault in 15 of
15 safety-critical situations. Three independently developed models spanning a
thirty-fold cost range exceeded the baseline. In a fully auditable evaluation,
these results demonstrate run-time abnormal situation management without a
human in the loop.
\end{abstract}

\section*{Introduction}

Autonomous systems now match or exceed expert human performance inside their
design envelope. Learned controllers shape tokamak
plasmas~\cite{degrave2022tokamak}, navigate stratospheric
balloons~\cite{bellemare2020balloon} and race quadrotors against world
champions~\cite{kaufmann2023drone}, and language-model agents plan and execute
chemical experiments~\cite{boiko2023coscientist}. What remains human is the
abnormal case. Bainbridge identified the division four decades ago, observing
that automation assumes the tasks that can be specified and leaves the operator
responsible for the abnormal conditions that cannot~\cite{bainbridge1983}. The
systems-and-control research agenda still describes deployed safety-critical
automation in the same terms: a system that operates itself within its design
conditions while a human supervises it and intervenes when its behaviour is no
longer appropriate or safe~\cite{lamnabhi2017agenda}. The division is codified
in autonomy taxonomies. In driving, the transition to high automation is
defined by whether the system itself performs the fallback to a minimal-risk
condition~\cite{sae2021j3016}. In rail, unattended operation is certified
where the fallback is to bring the train to a safe halt~\cite{iec62290}, the
basis on which driverless metro lines operate at scale. Where automation does
act outside its design envelope, it is through protective logic engineered or
trained offline for a specific anticipated
event~\cite{iec61511,seo2024tearing}. An abnormal situation in a chemical
plant, however, cannot be resolved by halting, because shutdown forfeits
production and subjects the plant to the startup and shutdown transients in
which process-safety incidents concentrate~\cite{ccps2021transient}. The
abnormal case therefore remains with the control-room operator, and chemical
plants remain the unsolved instance of this general boundary.

Modern chemical plants are operated through a layered control hierarchy.
Regulatory feedback loops, tuned for a nominal operating regime, hold the
process at its operating point. Above them, model predictive control
and, above that, real-time optimization coordinate the loops toward economic
objectives, with constraint handling that holds only while the process model
remains valid. Every layer of this hierarchy presupposes that the process stays
within a design envelope. Abnormal situations carry the process beyond that envelope,
where automated response is limited to alarm annunciation and to
safety-instrumented systems that trip the plant, bringing it to a safe state as
a last resort~\cite{iec61511}. These situations include equipment degradation,
off-design operation, sensor or actuator faults, and disturbances outside the
design basis.

The function that fills this gap is termed abnormal situation management. It calls for
detecting the departure, diagnosing its cause from incomplete and mutually
inconsistent measurements, deciding whether and how to correct it, and
initiating an orderly shutdown only when no corrective action can maintain
operation within constraints. The task is performed manually by control-room
operators, and mismanaged abnormal situations are a leading contributor to
process-safety incidents~\cite{nimmo1995,shu2016asm}. It has resisted automation for a
structural reason. Regulation reduces to a feedback law with provable guarantees. Abnormal
situation management comprises a sequence of diagnosis and decision problems
over heterogeneous evidence.

Large language models, and the large reasoning models (LRMs) that allocate
additional inference-time computation to intermediate reasoning, have been
applied to control predominantly outside the feedback loop. At design time,
they are used for reward synthesis~\cite{ma2023eureka,yu2023language2rewards},
controller tuning, controller-code drafting and automated control
design~\cite{guo2024controlagent,tohma2025smartcontrol,%
sun2024locomotion,narimani2025agenticcontrol}, and for translating
natural-language goals into
executable specifications~\cite{liang2023codeaspolicies,gkournelos2024hrc,%
lynch2021lcil,xiao2023skillaug}. At run time they are deployed in roles that
tolerate latency and occasional error, such as high-level planning above
conventional controllers~\cite{brohan2023rt1,brohan2023rt2,%
huang2022innermonologue,ahn2022saycan,xia2023modularproduction,cui2024survey}
and fault diagnosis without authority to
act~\cite{chiang2000,liu2024kgllm,%
alsaif2024multimodal,zhang2024buildingenergy,zheng2025kgfdd}. Attempts to place
a language model inside a classical control loop remain low-dimensional
feasibility demonstrations~\cite{maher2025llmpc}, in which an autoregressive
model provides neither a stability certificate nor bounded latency nor any
guarantee of constraint satisfaction.

A recent assessment identifies process control as the least mature use of
language models in process systems engineering and calls for
propose--validate--execute architectures, representative benchmarks and
failure-mode characterization~\cite{gopaluni2026}. Two requirements follow,
both addressed here. First, closed-loop performance under
disturbance has not been measured on a shared benchmark that permits methods
to be compared. Second, the scope of a model's authority over a physical
process, and the verification applied to its outputs, have not been specified
precisely enough to be quantified. The model-plus-check pattern is
recommended in the literature~\cite{sarhadi2025,gopaluni2026} and used in
knowledge-assisted diagnosis~\cite{liu2024kgllm}, but how much of the achieved
safety is contributed by the model and how much by the check has not been
quantified. Both the recommended pattern and the exclusion of language models
from authority rest on the same assessment: that language models are too
stochastic to be relied upon in or above a safety-critical
loop~\cite{sarhadi2025}.

Here we ask whether a large reasoning model can perform abnormal situation
management on a plant-wide process at run time, and under what conditions it
can do so reliably. We place the model in an abnormal-situation management
(ASM) layer above the regulatory control loops and constrain it to a bounded,
programmatically verified action interface. Within that interface it may adjust
regulatory setpoints within validated limits, activate pre-engineered
protective components, execute pre-audited procedures, take no action, or
initiate an orderly shutdown. Direct manipulation of actuators is not representable in the
interface. Every proposed action is checked against operating limits and, for
setpoint changes and reconfigurations, against a forward simulation.
The model is given only the information available to an operator. This
comprises a process description generated from the plant's instrumentation and
control configuration (its measurement tags and units, control loops, validated
operating limits and alarm settings), together with access to a nominal-model
forward simulator it can query to evaluate a candidate action.
The model receives no task-specific training and no fine-tuning, and is prompted
zero-shot, with no in-context demonstrations.

The interface the task requires is a contribution alongside the evaluation. We
built four components, a process-state summary generator that derives the
quantities an operator works from, the bounded action interface with its
programmatic validator, the forward simulation, and the two-part prompt. These
components present the state and bound the action space within which the model
makes its decisions.

We evaluate on the Tennessee Eastman process, an open-loop-unstable
benchmark derived from an industrial chemical process~\cite{downs1993}, in the
revised implementation of Bathelt, Ricker and Jelali~\cite{bathelt2015}, and
widely used for plant-wide control studies~\cite{mcavoy1994,larsson2001}. The
plant runs under a decentralized regulatory control layer designed for its
nominal operating envelope~\cite{ricker1996}; this baseline is the
layer above which the ASM layer acts. The comparison standard is a reference
controller that extends the baseline with two advanced regulatory control (ARC)
components from Ricker's published strategy~\cite{ricker1996}: composition
trims that maintain product grade, and override loops that shed production as a
constraint approaches its limit. The model is given neither the reference logic
nor the process knowledge behind it.

The benchmark subjects the plant, running under the baseline, to three
classes of scenario. The first requires no intervention; in the second, a
disturbance drives the baseline-controlled plant to a protective interlock
trip; in the third, a product-grade setpoint change exceeds what the baseline
can track. The question is whether a reasoning model can close the latter two gaps
without degrading performance on the first.

We find that it can. The reasoning model maintains the plant within all hard
constraints on every scenario that trips the baseline, and it remains within
constraints on every no-intervention scenario. Four
configurations of three independently developed reasoning models exceed the
baseline under the same architecture, and the primary endpoint separates none of them from the
reference.
In a matched case holding all else fixed, the model's reasoning effort alone
changed the outcome on the most demanding safety-gap scenario.

These results show that decision-making outside the process's design envelope,
the function that autonomy frameworks across domains assign to a human or to a
safe halt~\cite{bainbridge1983,sae2021j3016,iec62290,lamnabhi2017agenda}, can
be performed at run time by a general-purpose reasoning model whose authority
is bounded and verified. They quantify
how much authority can be
delegated to a language model over a physical process and under what
verification, and they provide a closed-loop benchmark and evaluation protocol
where the field has had none. More broadly, the conditions under
which authority can be delegated safely bear on autonomous plant
operation~\cite{gamer2020autonomous}.

\section*{Results}

\subsection*{A reasoning model as an abnormal-situation management layer above
basic regulatory control}

\begin{figure}[!tp]
\centering
\includegraphics[width=\linewidth]{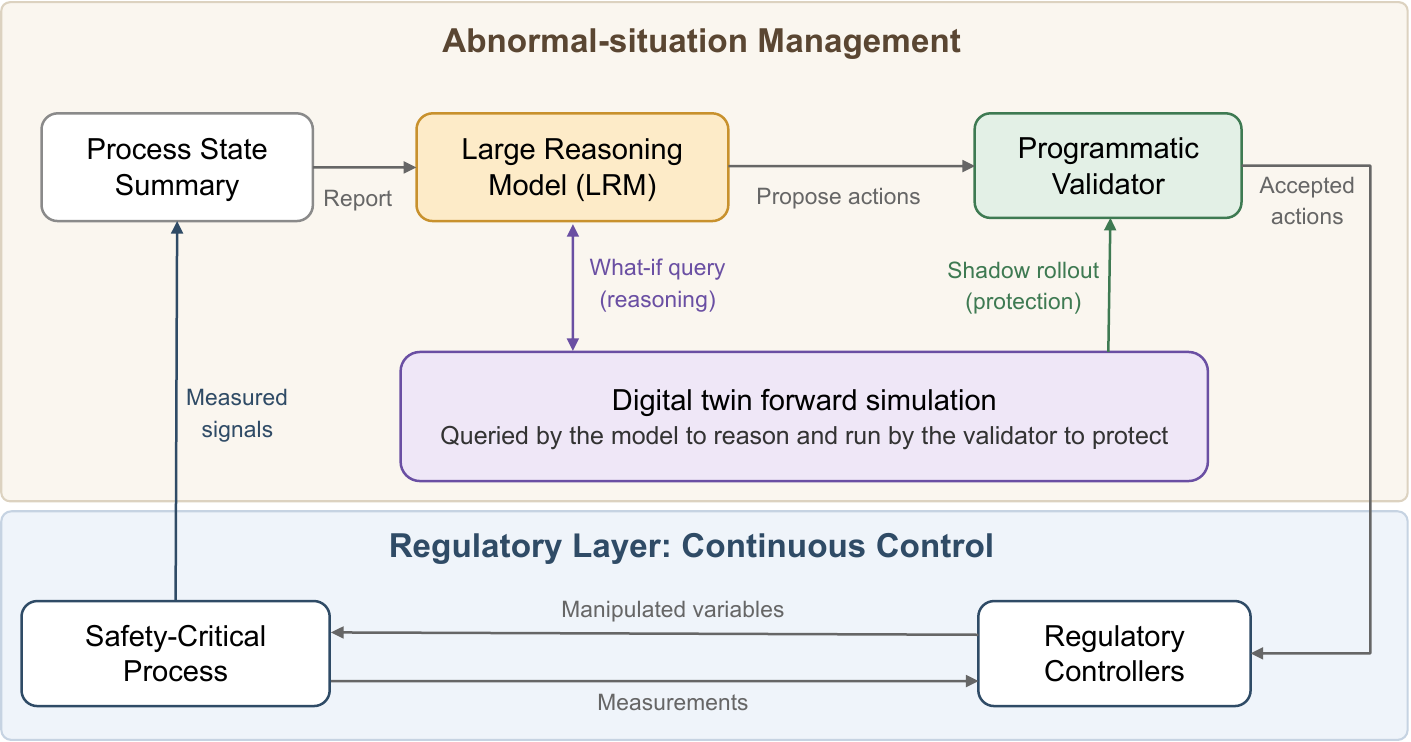}
\caption{\textbf{Abnormal-situation management layer and the bounded action
interface.}
A large reasoning model operates as an abnormal-situation management (ASM)
layer above a decentralized
proportional--integral regulatory layer running the Tennessee Eastman
process. The model is invoked on alarm onset (with a minimum inter-invocation
interval, a periodic check and a per-episode budget) and receives a process-state summary
constructed only from measured signals (\cref{fig:interface}). It returns one
diagnosis and one proposed action from the bounded interface; direct actuator
manipulation is not representable. A programmatic validator checks every proposal against ranges,
rates, cooldowns and budget, and returns machine-readable reasons for a single
revision. A single nominal-model forward simulation, with no access to the
true disturbance, serves two roles. The model queries it to evaluate a
candidate action before proposing (reasoning), and the validator runs it as a
shadow rollout to veto an action predicted to trip the plant (protection), so the
same simulation supports both the model's reasoning and the architecture's
protection layer.}
\label{fig:architecture}
\end{figure}

Within its nominal envelope the regulatory control layer requires no
supervision (\cref{fig:architecture}). Outside it, the layer has no provision
for correction, and a sufficiently severe disturbance drives the plant to a
protective interlock. The reference controller closes this gap with the two
engineered ARC components. The reasoning model, forming the ASM layer, instead
acts on the unmodified baseline. It is invoked on alarm onset, subject to a minimum inter-invocation
interval and a per-episode invocation budget, and on a periodic check. Accepted actions are applied after a modelled decision latency. This
follows the decomposition of slow supervisory reasoning from fast regulatory
control used in autonomous driving~\cite{sha2023languagempc}.

\begin{figure}[!tp]
\centering
\includegraphics[width=\linewidth]{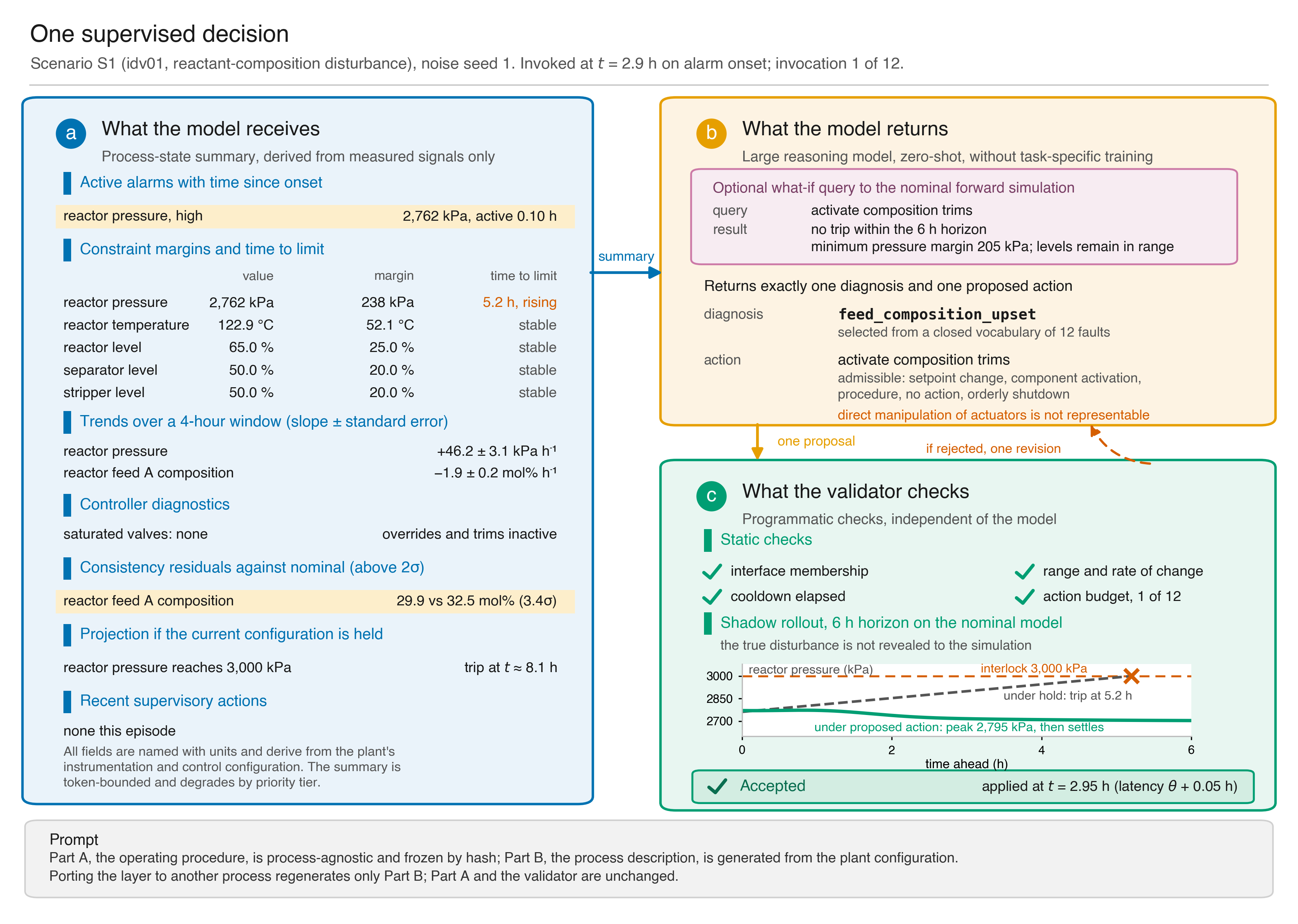}
\caption{\textbf{One supervised decision.} A single invocation of the
abnormal-situation management layer on safety-gap scenario S1 (kernel
disturbance idv01, a reactant-composition disturbance), noise seed 1, triggered
at $t = 2.9$~h by the reactor-pressure alarm (invocation 1 of 12).
\textbf{a}, The process-state summary the model receives, derived only from the
measured signals. It lists the active alarm, each constraint margin with its
signed time-to-limit, windowed trend slopes with standard errors, controller
diagnostics, consistency residuals against the nominal model, and a projection
of each constrained variable if the current configuration is held. Every field
is named and carries units from the plant configuration, and the summary is
token-bounded.
\textbf{b}, The reasoning model, prompted zero-shot, may first query the nominal
forward simulation with a candidate action (here activating the composition
trims) and returns exactly one diagnosis, from a closed twelve-item vocabulary,
and one action, from the bounded interface.
\textbf{c}, The programmatic validator, independent of the model, applies static
checks (interface membership, range and rate, cooldown, and the per-episode
action budget) and a 6-h shadow rollout on the nominal model, which is not given
the true disturbance, comparing the proposed action against holding. The
accepted action is applied at $t = 2.95$~h, after the modelled decision latency.
The prompt separates a process-agnostic Part~A, frozen by hash, from a Part~B
generated from the plant configuration.}
\label{fig:interface}
\end{figure}

The ASM interface converts the plant's measured variables into the
derived quantities an operator works from (\cref{fig:interface}). Before each model call, the ASM software computes
constraint margins with signed time-to-limit, trend slopes, controller
diagnostics, projections, and consistency residuals (Methods). These derived quantities form the
process-state summary the model receives in place of the raw measurement
vector. The
model returns a diagnosis from a predefined diagnostic vocabulary and one
action from the bounded interface. A programmatic validator gates every
proposal against static limits and, for setpoint changes and reconfigurations,
against a shadow forward simulation over a horizon that includes the inventory
transients a reconfiguration induces. On rejection it returns machine-readable
reasons and permits a single revision. The prompt is composed of a
process-agnostic operating procedure and a process description generated from
the same plant configuration, so applying the layer to a different process
requires regenerating only the description. The reference controller is run under the identical disturbance and noise
seed (Methods).

\subsection*{A benchmark of scenarios that leave the nominal envelope}

Abnormal situation management makes three distinct demands, and the
benchmark comprises 13 scenarios in three classes that test each demand
separately. The ASM layer must withhold intervention when the regulatory layer
suffices. Five no-intervention scenarios apply disturbances under which the baseline
remains within all constraints. The ASM layer must correct a trajectory toward
a safety limit. In
5 safety-gap scenarios the baseline reaches the reactor-pressure interlock on
every seed, while the reference controller maintains operation, confirming that
the episodes are recoverable. The ASM layer must restore commanded performance.
In 3 quality-gap scenarios a product-grade setpoint change exceeds what the
baseline can track. The primary endpoint is whether the plant is stabilized within all hard constraints at the horizon, with an interlock trip
and an orderly shutdown both counting as failure.

Four properties make the suite a controlled benchmark. The process is perturbed
exclusively through the native disturbance channels and exposed setpoints of the
public simulation kernel of Bathelt et al.~\cite{bathelt2015}, with no kernel
modification and no synthetic faults. Scenario severities were calibrated
empirically before any language-model experiment so that the class-defining
behaviour holds on every seed, and the suite is frozen and
never used for prompt iteration. Three paired noise seeds per scenario fix the
measurement-noise realization across configurations, so every comparison is
paired. Every decision transcript is cached, making the campaign exactly
replayable offline.

\subsection*{The reasoning model maintains the plant within constraints on
scenarios that trip basic regulatory control}

Under the baseline the plant is stabilized within all hard constraints in 24 of
39 episodes, tripping in all 15 safety-gap episodes. On the quality-gap class it
remains within hard constraints but cannot track the product-grade change, and
on the no-intervention class no action is required. Claude Sonnet-5 at high
reasoning effort stabilizes the plant within all hard constraints in all
39 episodes, recovering every safety-gap episode and meeting the endpoint
on every no-intervention episode (\cref{tab:headline}). This is a paired
difference in the in-constraint fraction of
$+0.385$ over the baseline (95\% confidence interval $[+0.22, +0.54]$,
Newcombe's method for paired proportions; exact McNemar $p = 6\times10^{-5}$,
$n = 39$), concentrated in the safety-gap class, where the difference is
$+1.00$. On the primary endpoint the reference controller reaches the same 39
of 39, with no discordant pairs between the model and the reference (0 of 39;
95\% CI on the difference $[-0.09, +0.09]$), so the primary endpoint does not
distinguish them. Once the process-state summary reports the operating-target
deviation, the model tracks the commanded product-grade change on all 3
quality-gap scenarios (\cref{tab:summary}).

\begin{table}[t]
\centering
\caption{\textbf{Outcome summary by configuration} (13 scenarios $\times$ 3
paired noise seeds = 39 episodes per configuration). In-constraint fraction is the
fraction of episodes in which the plant is stabilized within all hard constraints
at the horizon (interlock trip and orderly shutdown both count as failure; no
model-supervised episode ended in an orderly shutdown). Diagnosis accuracy (any
invocation) is the agreement of the reported diagnosis with the ground-truth tag
on any invocation of an episode; this is a best-of-$n$ metric (Methods). Cost is
the model cost per episode at provider list prices in July 2026.}
\label{tab:headline}
\small
\begin{tabular}{lccccc}
\toprule
Configuration & In-constraint & Trips & \shortstack{Diagnosis acc.\\(any inv.)} & \shortstack{Actions/\\episode} & Cost (US\$) \\
\midrule
Baseline & 0.615 & 15 & -- & -- & -- \\
Reference controller & 1.000 & 0 & -- & -- & -- \\
\addlinespace
Sonnet-5, high effort (primary) & 1.000 & 0 & 0.846 & 2.54 & 2.32 \\
Sonnet-5, low effort & 0.974 & 1 & 0.821 & 2.82 & 1.31 \\
GLM-5.2 & 0.949 & 2 & 0.769 & 3.64 & 0.67 \\
DeepSeek-V4-Flash & 1.000 & 0 & 0.744 & 1.85 & 0.07 \\
\bottomrule
\end{tabular}
\end{table}

\subsection*{The result reproduces across models and cost, and reasoning
effort separates a matched pair}

Repeating the campaign with Sonnet-5 at low reasoning effort and with two
independently developed reasoning models, GLM-5.2 and DeepSeek-V4-Flash, under
the same architecture, prompt and interface yields the same conclusion
(\cref{tab:headline}): DeepSeek-V4-Flash remains in constraint
in all 39 episodes; Sonnet-5 at low effort in 38; GLM-5.2 in 37. Each
configuration exceeds the
baseline, with paired differences in the in-constraint fraction of $+0.33$ to
$+0.39$ (exact McNemar $p = 6\times10^{-5}$ to $2\times10^{-3}$); none is
distinguishable from the reference on the primary endpoint (Newcombe 95\%
confidence intervals include zero). The configurations are not distinguishable
from one another
at this sample size (pairwise McNemar $p \ge 0.5$). Costs span more than a
thirty-fold range, approximately US\$2.80 to US\$90.48 per 39-episode campaign
(\cref{tab:headline}). The least expensive
configuration attains the same
in-constraint fraction as the most expensive, so the outcome is not a property
of a single model.

\begin{figure}[!tp]
\centering
\includegraphics[width=0.9\linewidth]{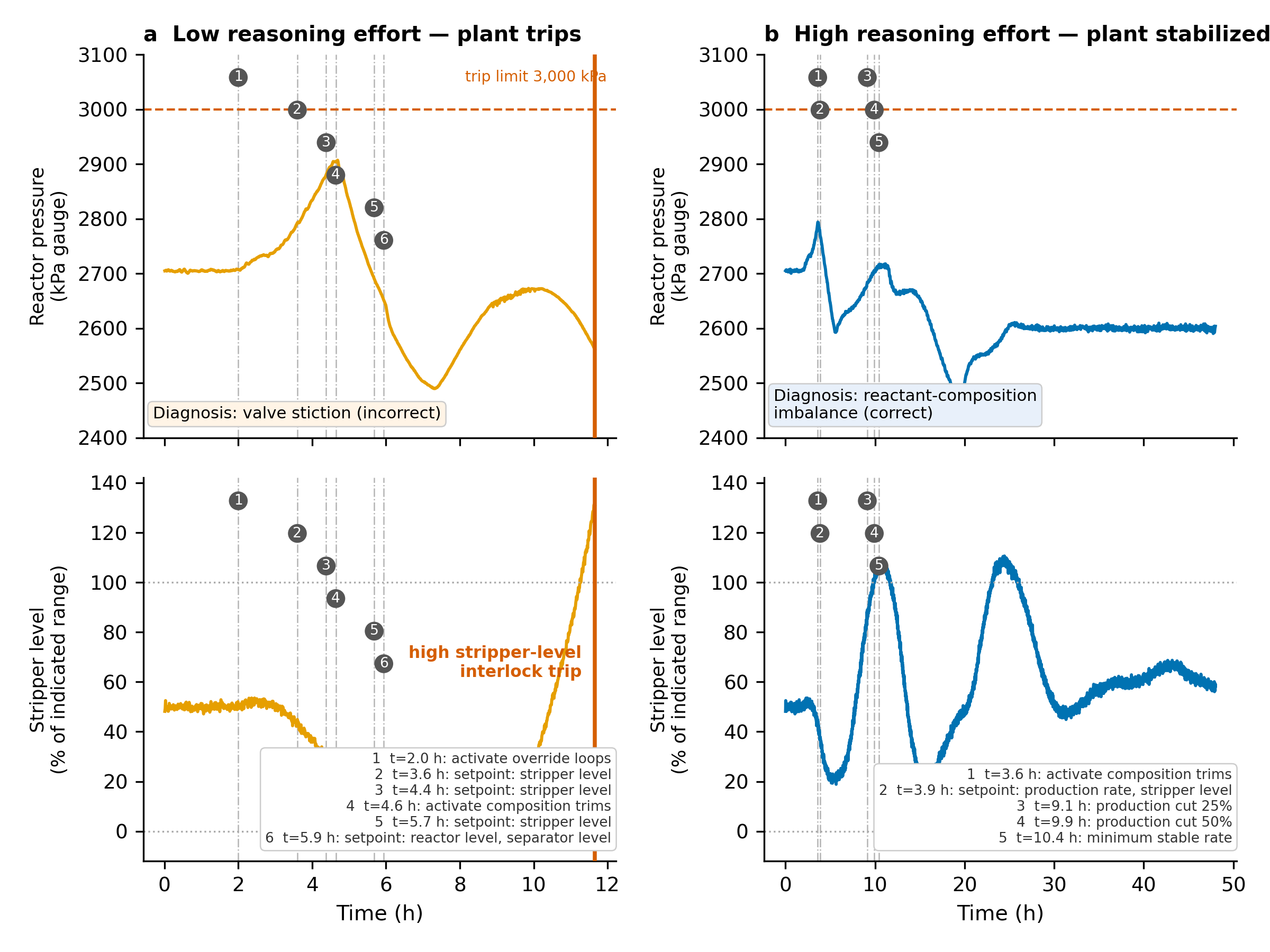}
\caption{\textbf{Reasoning effort separates a trip from stabilization on the
most demanding scenario.} Reactor pressure (top) and stripper level (bottom) on the identical
scenario and noise seed, a reactant-composition disturbance introduced at
elevated production, for Claude Sonnet-5 at low (\textbf{a}) and high
(\textbf{b}) reasoning effort, with the model's accepted actions numbered on
the trajectories. At low effort the model diagnoses valve stiction (incorrect) and the stripper
inventory reaches the high-level interlock at $t = 11.6$~h. At high effort the
same model diagnoses the reactant-composition imbalance (correct), activates
the composition trims with staged production cuts, and the plant is stabilized
for the full 48-h horizon. The dotted lines at 0\% and 100\%
mark the level instrument's indicated range. The interlock threshold acts on
holdup volume. Values plotted beyond this range are linear extrapolations of
the indicated signal. Note the different time axes, because the
low-effort episode ends at the trip.}
\label{fig:effort}
\end{figure}

In the one failed safety-gap episode, the depth of reasoning changed the
outcome. The episode is
informative because a matched comparison isolates the cause. On the most demanding
safety-gap scenario (S4), a reactant-composition disturbance at elevated
production, Sonnet-5 at low reasoning effort tripped the plant, whereas the
same model at high effort, on the identical scenario and noise seed, stabilized
it (\cref{fig:effort}). At low effort the model identified the cause as
valve stiction, activated the override loops and issued repeated
setpoint corrections. The disturbance persisted and the stripper inventory rose
to its high-level interlock. At high effort the model identified the
reactant-composition
imbalance, activated the composition trims and, anticipating the inventory
transient the activation induces, staged production cuts that held the vessel
levels within limits while the trims took effect. The difference lies in
diagnosis and the choice of protective component. Both configurations issued
admissible, validated actions throughout. The effect is not statistically
resolvable at this sample size (one discordant pair over 39; McNemar
$p = 1.0$), and we report it as a single case with an identified mechanism, without
estimating an effect size.

\subsection*{Diagnosis distinguishes the model from a rule table}

\begin{figure}[!tp]
\centering
\includegraphics[width=0.82\linewidth]{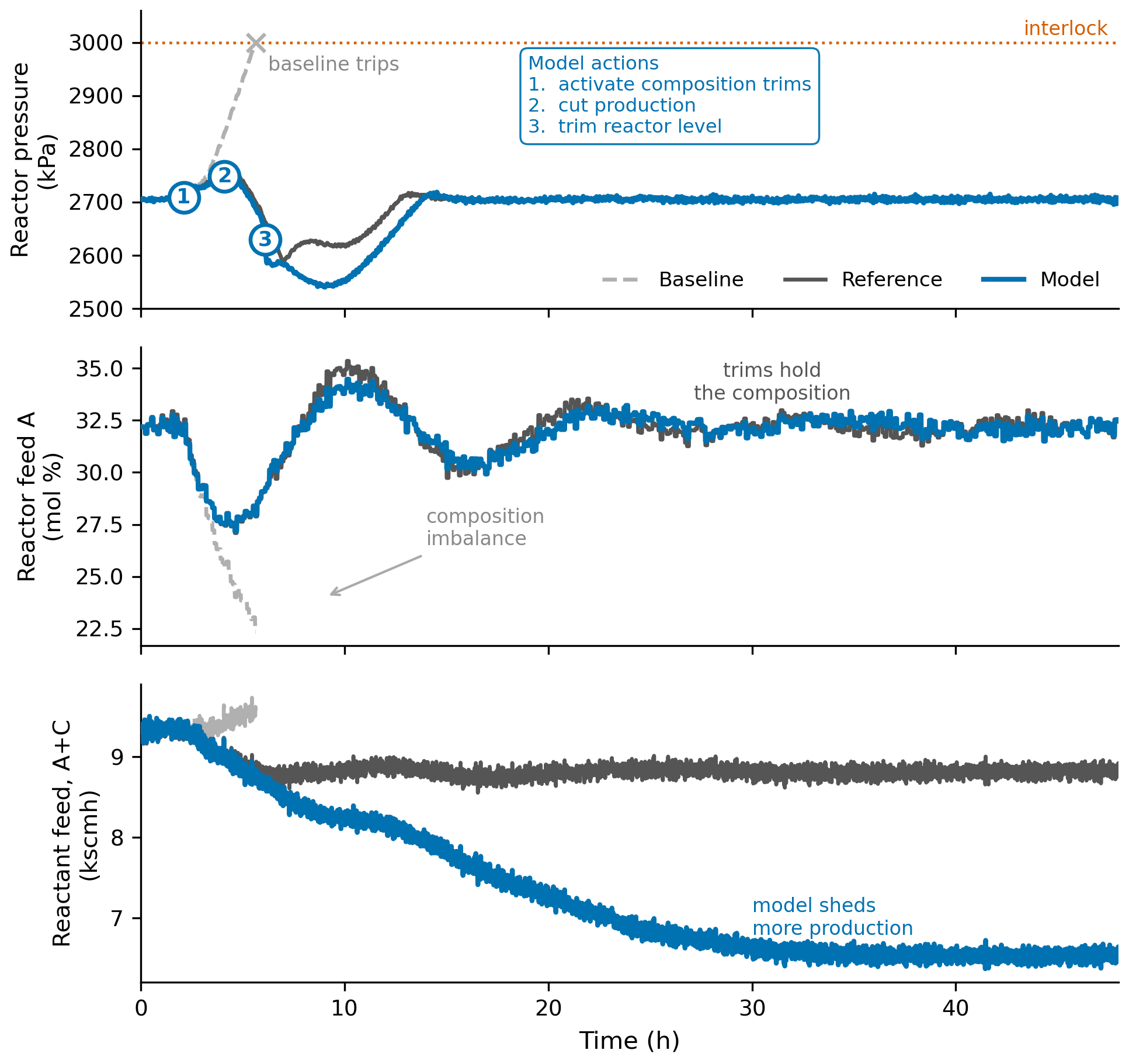}
\caption{\textbf{One safety-gap episode.} Reactor pressure (top), reactor feed A
composition (middle) and reactant feed (bottom, in thousand standard cubic
metres per hour) for the baseline, the reference
and the model on a reactant-composition disturbance (S1). \textbf{Top}, the
baseline rises to the reactor-pressure interlock and trips at $t \approx 6$~h,
whereas the reference and the model remain below it for the full 48-h
horizon; the model's accepted actions are numbered on its trajectory, activating
the composition trims (1), cutting production (2) and trimming the reactor level
(3). \textbf{Middle}, the disturbance drives the reactor feed A composition away
from its operating value; the baseline allows it to diverge, while the composition
trims restore it under the reference and the model. \textbf{Bottom}, reactant
feed, which the model reduces further than the reference.}
\label{fig:mechtraj}
\end{figure}

How the model maintains the plant within constraints is visible in the actions
it takes. On the safety-gap class the model averaged approximately four accepted
actions per episode. One episode illustrates the sequence. The model diagnoses the composition
upset, takes the three actions numbered in \cref{fig:mechtraj}, and maintains
operation while the baseline trips. It applies
these actions more conservatively than the reference, holding a slightly larger
margin below the interlock (mean peak pressure 2{,}774~kPa versus 2{,}797~kPa;
$n = 15$ paired episodes) while running the reactant feed approximately 19\% below the
reference, a paired mean difference of 1.8 thousand standard cubic metres per
hour (kscmh; 95\% confidence interval $[1.5, 2.3]$). This reduction reflects a default that prioritizes safety margin
over throughput, consistent with a layer whose function is to prevent trips. The reference, tuned for this plant's economics, operates closer to the
constraint.

The same episodes show the model diagnosing the fault before selecting its
response. It identifies the ground-truth tag on every safety-gap episode
(15/15). A scripted rule table reacts to alarm thresholds without forming a
diagnosis. Given the identical process-state summary, interface and validator,
it reaches the same in-constraint outcome but identifies the cause in only 2 of
15 (Supplementary Note~5). The endpoint is reachable by a fixed rule table. The
diagnosis and the choice of protective component set the model apart.

\subsection*{The process-state summary determines which conditions the model
can act on}

The process-state summary converts the plant's measured signals into the
derived quantities the model reasons from, and its content sets which
conditions the model can act on. For every hard limit the summary reports the
margin to the limit and the projected time to reach it, but it presents the
operating targets only as setpoints beside their measurements. The effect is
measurable on the quality-gap class. The model recognizes the unmet product
grade and activates the composition trims on 6 of the 9 quality episodes and,
on those, tracks the commanded grade to within 0.4~mol\%. On the grade-reduction
scenario with no concurrent disturbance, it takes no action on any of the three
seeds, and the grade settles 13.6~mol\% from its commanded value, no better
than the baseline.

Reporting one further derived quantity removes the asymmetry. When the summary
states each commanded operating target against its delivered value as a
deviation from setpoint, in the same form as the constraint margins, the model
reads the shortfall, diagnoses the operating-point change, and activates the
composition trims with a staged production cut for the induced transient. It
then stabilizes the grade on all 9 quality episodes and reduces the mean
steady-state grade offset from 4.8 to
0.3~mol\%, matching the reference controller, in constraint throughout
(\cref{tab:summary}). The safety-gap and no-intervention outcomes are
unchanged, because their operating targets are met and the added quantity
reports no deviation. The limiting factor is the content of the summary.

\begin{table}[t]
\centering
\caption{\textbf{Reporting the operating-target deviation allows the model to track
the commanded product grade} (primary model, quality-gap class, three paired
seeds per scenario). With the deviation reported in the process-state summary
the model activates the composition trims and drives the product grade to its
setpoint on every scenario; without it, the model leaves the grade-reduction
scenario uncorrected.}
\label{tab:summary}
\small
\begin{tabular}{lcccc}
\toprule
& \multicolumn{2}{c}{Deviation not reported} & \multicolumn{2}{c}{Deviation reported} \\
\cmidrule(lr){2-3}\cmidrule(lr){4-5}
Scenario & Trims activated & Grade offset & Trims activated & Grade offset \\
 & (of 3) & (mol\%) & (of 3) & (mol\%) \\
\midrule
Grade $54 \to 40$ & 0 & 13.6 & 3 & 0.2 \\
Grade $54 \to 65$ & 3 & 0.4 & 3 & 0.4 \\
Grade $54 \to 40$, disturbance & 3 & 0.2 & 3 & 0.2 \\
\addlinespace
All (9 episodes) & 6 & 4.8 & 9 & 0.3 \\
\bottomrule
\end{tabular}
\end{table}

\subsection*{The failures are few, and no single mechanism recurs}

\begin{figure}[!tp]
\centering
\includegraphics[width=0.9\linewidth]{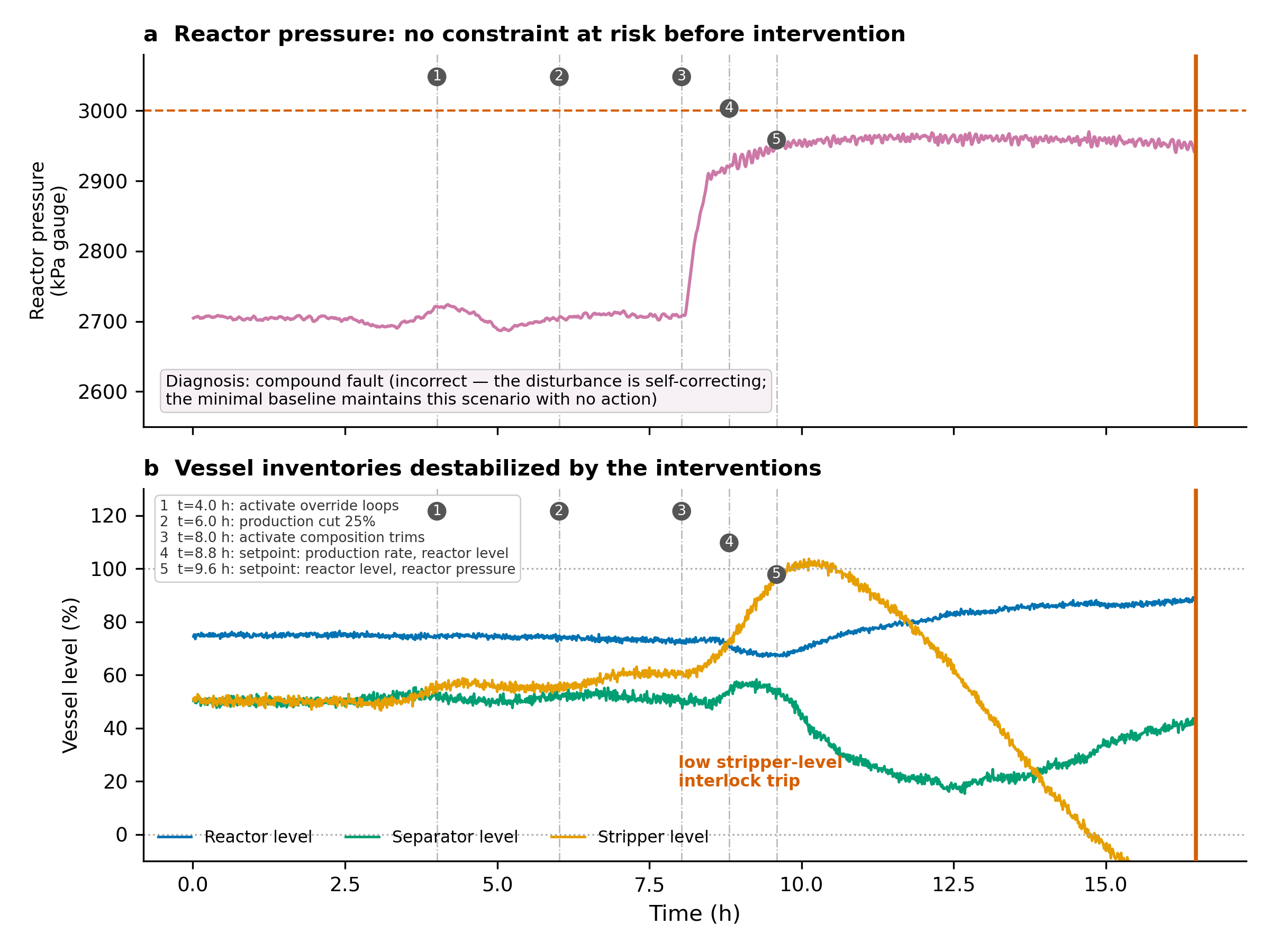}
\caption{\textbf{The second failure mechanism, intervention on a
self-correcting disturbance.} A GLM-5.2 episode on a random feed-composition
disturbance that the baseline maintains with no action. \textbf{a},
Reactor pressure is at its nominal value with no constraint approaching a limit
when the model classifies the condition as a compound fault and begins
intervening (accepted actions numbered). \textbf{b}, The interventions
(activating the override loops, cutting production, activating the composition
trims and moving setpoints) destabilize the vessel inventories; the stripper
drains to the low-level interlock at $t = 16.5$~h. The correct decision
throughout was to withhold action. This is one of the three failures across
156 model-supervised episodes and the only one on a scenario requiring no
correction; the other two are a diagnostic error under reduced reasoning
(\cref{fig:effort}) and a diagnostic revision on a scenario that did require
correction (Supplementary Note~11).}
\label{fig:overintervention}
\end{figure}

Three of the 156 model-supervised episodes ended in a trip, each by a distinct
mechanism. The first is
the low-effort misdiagnosis above, in which valve stiction was identified in
place of a reactant-composition imbalance. The second is on a scenario that
requires no correction, where GLM-5.2 classified a self-correcting random
feed-composition disturbance as a compound fault and intervened. The
interventions drove the vessel inventories to a level interlock (\cref{fig:overintervention}).
The third is on a scenario that does require correction, where GLM-5.2 first
identified a composition upset correctly and activated the composition trims,
then revised its diagnosis to a feed-loss condition that was not present. The
setpoint corrections issued under the revised diagnosis drove the
stripper inventory to its high-level interlock (Supplementary Note~11). No
failure involved an action that the validator had rejected, and each was
confined to the single scenario in which it occurred. The three mechanisms are
diagnostic error under reduced reasoning, unnecessary intervention, and
revision away from a correct initial diagnosis.

The model does not always withhold action on scenarios requiring no correction.
Across the four configurations, in 42 of 60 such episodes an intervention was
proposed, validated and applied although none was required. Only the episode
described above proved consequential (Supplementary Table~S6). The low
failure rate on this class therefore shows that unnecessary action is
usually inconsequential (41 of 42 episodes).

The reliable decision-making reported above could in principle be an artefact
of the verification layer, with the validator
rejecting unsafe proposals until an admissible one passes. A verification
ablation rules this out. We re-ran the primary configuration on the identical
scenarios and paired seeds with the run-time checks progressively removed:
first without the forward-simulation veto, then with no validator at all. Both
configurations maintain the plant within all hard constraints in 39 of 39
episodes. In the three
episodes where the validator had rejected a proposal, the no-validator
configuration applied the original proposal and still reached the same outcome
(Supplementary Note~10; Methods). An audit of the primary campaign shows the
veto never fired and the static checks rejected three proposals, all three at or
marginally beyond static limits. With one
qualification (the three statically rejected proposals were revised), the
decisions reported here are the model's own. In this campaign, verification
activity differed across configurations (3, 10, 13 and 27 static
rejections for Sonnet-5 at high effort, Sonnet-5 at low effort,
DeepSeek-V4-Flash and GLM-5.2 respectively).

\section*{Discussion}

We have shown that a large reasoning model performs abnormal situation
management on a plant-wide process at run time, with its authority confined to
the bounded, verified action interface above basic regulatory control. Across abnormal scenarios that reliably trip
the baseline, four configurations of three reasoning models exceed the
baseline, and none is distinguishable from the process's
advanced regulatory control on the primary endpoint.

Three implications follow for the use of reasoning models in safety-critical
control. First, a general-purpose reasoning model can make correct
decisions on a plant-wide process. The results reported here contradict
the assessment that language models are too stochastic to be relied upon in or
above a safety-critical loop. The verification ablation is consistent with these
decisions being the model's own: removing the checks changed no primary-endpoint
outcome (Supplementary Note~10). The bounded interface therefore fixes what an incorrect
decision can touch, and the programmatic verification is an independent
protection layer on which the performance does not depend. This is the
in-built verifiable monitoring mechanism that the systems-and-control research
agenda identified as a requirement for deployable
autonomy~\cite{lamnabhi2017agenda}. This
separation of function refines the model-plus-check pattern on which the
literature has converged~\cite{liu2024kgllm,sarhadi2025} and extends it from
design-time and open-loop uses to closed-loop operation.

Second, in a matched case reasoning effort separated the outcomes on the
most demanding scenario through diagnosis and intervention selection,
indicating where inference-time computation is best allocated in this setting. This finding extends the evidence for inference-time
computation~\cite{deepseek2025r1,snell2024scaling}, including the use of
simulators as reasoning environments~\cite{prabhudesai2026sim2reason}, from
textual and mathematical benchmarks to a metric of physical safety.

Third, that four configurations across a thirty-fold cost range are
indistinguishable from the reference on the primary endpoint shows that the
result rests on the surrounding system. The summary generator, the bounded interface and its verification, and the
two-part prompt make a general model reliable here. The model is a replaceable
component. The recurring inference cost of
a layer of this kind can therefore be low, though the non-recurring engineering
cost of the surrounding system is not addressed here. The architecture, a conventional inner
regulatory loop with a validated reasoning layer above it, is not specific to
this process, and language-model decision layers are being investigated
across
process operations, industrial digital twins and adjacent decision-making
domains~\cite{da2024prompt2transfer,yang2024llmdt,%
huang2025roboticsynthesis,fan2026maintenance,lao2024gptuner,%
decardinelson2024genai}. The protocol supports comparing architectures, action
interfaces and verification strategies as well as models.

Several limitations bound these claims. The evaluation is in simulation. The
sim-to-real gap for language-model control of this kind has not been quantified
in any published study, and is not addressed here. The
layer's forward-simulation interface is a nominal model. The nominal forward simulation can predict as safe an action
that later contributes to a trip, because it is not given the true disturbance.
This occurred in the longer-span failures
(\cref{fig:overintervention}; Supplementary Note~11). That is the
expected behaviour of an operator's simulator, and a residual failure mode that
would persist in deployment. Divergence between plant behaviour and the layer's
nominal model is the regime in which a human operator's familiarity with the
specific plant is most valuable. We compare against engineered protective logic throughout. We make no
comparison with human operators.

The run-time verification is likewise bounded in what it can detect. It
evaluates one proposed action at a time over a fixed horizon. The failure
on a scenario requiring no correction and the diagnostic revision on a scenario
that required one both developed over longer spans,
through sequences of individually admissible actions that the
per-action check did not flag. Verification at the level of the evolving trajectory is the corresponding
direction for this layer.

The safety-gap class is solvable on this benchmark by activating components
that the baseline leaves disabled, as the scripted-supervisor comparison
showed. The outcome-level result therefore demonstrates that the
architecture, given a library of pre-engineered protective components, maintains
the plant within constraints. The demonstrated contribution of reasoning lies
in diagnosis accuracy. Accepted actions per episode do not track diagnosis
accuracy across configurations (\cref{tab:headline}), and we make no parsimony
claim.
Abnormal situations that no pre-engineered component or procedure addresses are
not represented and would be a harder test of the layer's reasoning. With the
in-constraint fraction at or near 1.00 for every configuration, the
protocol separates the reasoning configurations from the baseline on the primary
endpoint, and from the non-reasoning comparator on diagnosis accuracy. It does
not separate the configurations from one another on either endpoint.

No formal guarantee or safety-integrity-level claim is made. The interface and
validator bound the
consequences of an incorrect decision and make failures inspectable, which is
the standard for a decision layer that acts below the safety-instrumented
system and leaves it in place. The evaluation is on a single process.
Generalization is supported in principle by the separation of the
process-agnostic operating procedure from the generated process
description, but is not demonstrated here. The cost comparison spans list
prices across different providers and inference services, and is not a
controlled cost-capability frontier.

The result nonetheless changes the status of abnormal situation
management as an automation target. The function has required either a human
operator or protective logic engineered offline for anticipated cases. A
general-purpose reasoning model now performs it at run time, under conditions
stated in a form that can be measured and compared, on a benchmark that others
can run. They are reasoning sufficient for the diagnosis, an interface that
bounds the scope of an error, and an independent verification layer retained as protection.
Several next steps follow. A second process family would test the generality
that the process-agnostic operating procedure is intended to provide. Scenario
families hard enough to
separate the model configurations would test distinctions the present benchmark
cannot resolve. The safety--economics balance is set by the operating
procedure, and the model currently sheds more production than the plant-tuned
reference requires. A clause that restores throughput once safety is secured
would address this default.

\section*{Methods}

\subsection*{Formal statement of the abnormal-situation management problem}

We adopt the primary- and secondary-endpoint convention of clinical trials.

\textbf{Plant and constraints.} The plant state $x(t) \in \mathbb{R}^{50}$
evolves under fixed-step numerical integration of the kernel dynamics
\begin{equation*}
\dot{x}(t) = f\bigl(x(t),\, u(t),\, d(t)\bigr), \qquad
y(t_k) = h\bigl(x(t_k)\bigr) + v_k,
\end{equation*}
where $u \in \mathbb{R}^{12}$ are the manipulated variables, $d(t) \in
[0,1]^{28}$ the kernel's native disturbance channels (set by the scenario
schedule and never visible to any controller or the ASM layer), $y \in
\mathbb{R}^{41}$ the measured outputs, and $v_k$ measurement noise whose
realization is fixed by the episode seed. The interlock system halts the
plant at the stopping time
$\tau = \inf\{\,t : g(x(t)) > 0\,\}$,
where $g$ encodes the hard constraints (reactor pressure, reactor
temperature, and the three vessel-level limits), evaluated on the true state
$x$, independent of the measured outputs $y$.

\textbf{Regulatory layer.} At scan instants spaced $\Delta_c = 36$~s the
regulatory layer applies the decentralized proportional--integral law
$u(t_k) = \kappa\bigl(y(t_k),\, \xi_k;\, s,\, m\bigr)$,
where $\xi_k$ is the vector of controller integrator states,
parameterized by the setpoint vector $s \in \mathbb{R}^{9}$ and the
configuration $m = (m_{\mathrm{trim}}, m_{\mathrm{ovr}}) \in \{0,1\}^2$
(reactant-composition trims and constraint-override loops). The baseline
fixes $m = (0,0)$ and the reference fixes $m = (1,1)$. The
ASM layer may raise either component of $m$ at run time (activation is
one-way within an episode).

\textbf{Abnormal-situation management layer.} Invocation times
$\theta_1 < \theta_2 < \cdots$ are triggered by a new alarm or a periodic
check, subject to a minimum inter-invocation spacing and a per-episode
invocation budget. At
$\theta_i$ the ASM software computes the process-state summary
$R_i = \rho\bigl(y_{[\theta_i - W,\, \theta_i]},\, \mathcal{A}_i,\, s,\, m,\,
a_{1:i-1}\bigr)$
from the trailing measurement window ($W = 4$~h), the active alarm set
$\mathcal{A}_i$, the current configuration, and the layer's own past
actions (Supplementary Note~3 defines every field of $\rho$). The ASM layer
is a conditional distribution
\begin{equation*}
\pi\bigl(\hat{c}_i,\, a_i \mid P,\, R_i\bigr),
\end{equation*}
realized by sampling once from the language model under the frozen prompt $P$:
a diagnosis
$\hat{c}_i \in \mathcal{C}$ from the closed twelve-item vocabulary and one action
$a_i \in \mathcal{A}(s) = \mathcal{A}_{\mathrm{sp}} \cup
\mathcal{A}_{\mathrm{cfg}} \cup \mathcal{A}_{\mathrm{proc}} \cup
\{a_{\varnothing},\, a_{\mathrm{stop}}\}$:
a bounded setpoint change, a configuration activation, a pre-audited
procedure, no action, or an orderly shutdown. Direct assignment to $u$ is not
an element of $\mathcal{A}$. Before proposing, $\pi$ may query the what-if
map $\Sigma(a; \hat{x}_i) \mapsto$ (predicted constraint margins,
time-to-trip, product composition), the nominal model rolled forward from a
state estimate $\hat{x}_i$ formed from the measured outputs, with the
disturbance held at its last inferred value.

\textbf{Verification.} An accepted action must satisfy the predicate
$V(a_i \mid s, a_{1:i-1}) = V_{\mathrm{static}} \wedge V_{\mathrm{shadow}}$.
$V_{\mathrm{static}}$ checks membership, absolute range, per-move magnitude,
per-setpoint cooldown and the episode action budget.
$V_{\mathrm{shadow}}$ simulates the closed loop over a fixed horizon from the current state
estimate, once under $a_i$ and once under the hold (null) action
$a_{\varnothing}$, and
rejects if the action is predicted to trip when
holding does not, to trip more than $0.1$~h earlier than holding, or to
reduce a standing pressure or temperature margin below its floor by more than
the hold action does. On rejection the ASM layer may revise once. The verification-ablation configurations replace $V$ by
$V_{\mathrm{static}}$ (no-shadow) or by acceptance of every proposal
(no-validator).

\textbf{Endpoints.} With horizon $T = 48$~h, the primary endpoint of an
episode is
\begin{equation*}
S \;=\; \mathbf{1}\bigl[\tau > T\bigr] \cdot
\mathbf{1}\bigl[a_i \neq a_{\mathrm{stop}} \text{ for all } i\bigr],
\end{equation*}
the plant stabilized within all hard constraints. Diagnosis
correctness is the indicator
$D = \mathbf{1}\bigl[\exists\, i:\ \hat{c}_i \in \mathcal{C}^{*}\bigr]$
against the scenario's ground-truth tag set
$\mathcal{C}^{*} \subset \mathcal{C}$ ($|\mathcal{C}^{*}| \le 2$), fixed
before the campaign.

\subsection*{Process, control layers, and benchmark construction}

\textbf{Process.} The Tennessee Eastman process~\cite{downs1993} in
the revised implementation of Bathelt, Ricker and Jelali~\cite{bathelt2015}: a
plant-wide, open-loop-unstable process (reactor, condenser, vapour--liquid
separator, recycle compressor with purge, product stripper) with 41
measurements, 12 manipulated variables, and hard interlocks that shut the
process down at reactor pressure 3{,}000~kPa gauge, reactor temperature
175~\textdegree C, and vessel-level limits (reactor holdup outside
2--24~m$^3$, separator outside 1--12~m$^3$, and stripper outside 1--8~m$^3$,
equivalently the 0\% and 100\% bounds of each indicated level range), enforced
by the kernel on the true state. Simulations use a fixed integration step of
1.8~s with a regulatory scan of 36~s; episodes run to a 48-h horizon unless an
interlock trips or an orderly shutdown is ordered.

\textbf{Control layers.} The regulatory layer is Ricker's decentralized
proportional--integral strategy~\cite{ricker1996}. We build two instantiations
from the published strategy's own components: the \emph{baseline}
disables the reactant-composition trim loops and the constraint-override loops;
the \emph{reference controller} enables both and reproduces the published
strategy. The ASM layer may activate either component at run time, which is the
reconfiguration action class.

\textbf{Benchmark.} Thirteen scenarios in three classes perturb the process
exclusively through the kernel's native disturbance channels and exposed
setpoint schedules; the kernel is not modified and no synthetic sensor or
actuator faults are injected. Class~A (5 scenarios): disturbances under which the baseline
remains within constraints, including the idv06 feed loss at nominal production; the same channel
at maximum production is safety-gap scenario S5. The correct behaviour is to
withhold intervention. Class~S (5): (i) a reactant-composition imbalance alone; (ii) the
same combined with a second disturbance; (iii) at reduced severity; (iv) at
elevated production; and (v) a feed loss at maximum production; the baseline
reaches the reactor-pressure interlock in every episode and the reference
maintains operation. Class~Q (3): product-grade setpoint changes the baseline cannot track
(servo failure without the trims). We calibrated scenario severities
empirically before any language-model experiment: we ran each (scenario, seed)
pair under both the baseline and the reference controller, and class membership
required consistency on every seed (13/13 class-consistent). We freeze the suite
definition by content hash and never use the benchmark scenarios for prompt
iteration; development used a disjoint pilot set, after which the prompt was
frozen in the same way.

\textbf{Pairing.} The experimental unit is the (scenario, seed) episode. A
seed determines the kernel's measurement-noise realization, so all
configurations are subjected to identical conditions in every episode, and
every contrast between configurations is paired. The seed constrains only the measurement
noise.

\subsection*{Abnormal-situation management layer: process-state summary, action
interface, and validator}

The ASM layer comprises four engineered parts, a process-state summary
generator, a bounded action interface, a programmatic validator, and a forward
simulation, driven by a two-part prompt. We specify each below and in full in
Supplementary Notes 1 to 4.

\textbf{Invocation.} The ASM layer is invoked when a new alarm is annunciated
(annunciation limits are placed strictly inside the trip limits with hysteresis
and on-delay), subject to a minimum inter-invocation interval of 0.25~h, a
periodic check every 2~h, and a budget of 12 invocations per episode. An
accepted action is applied after a modelled decision latency of 0.05~h; the
plant is stepped deterministically, so provider latency does not enter
simulated time and is logged separately.

\textbf{Process-state summary.} Each invocation receives a token-bounded process-state summary constructed
only from measured signals: active alarms with ages; windowed least-squares
trend slopes with standard errors; controller diagnostics including saturated
valves and active overrides; constraint margins each with a signed
time-to-limit (margin over trend slope, reported as stable when the slope is
directed away from the limit, or when its magnitude is smaller than its standard
error); material-balance
and coupled-instrument consistency residuals derived from the plant schema,
reported only when they exceed $2\sigma$ of their rolling standard deviation
(window $W$); the
layer's own recent actions with a predicted-versus-observed convergence
flag; a projection of each constrained variable if the current configuration is
held (the nominal model rolled forward under no intervention); and a
comparison of the measured trajectory against the nominal model over the recent
window. Under alarm floods, reports truncate through a fixed sequence of
priority-ordered tiers. A further derived field, the deviation of
each commanded operating target (production rate and product grade) from its
delivered value, is evaluated on the quality-gap class (\cref{tab:summary}).

\textbf{Prompt.} The system prompt is composed of two parts frozen by content
hash. Part~A is a process-agnostic operating procedure: a lexicographic
objective ordering (process safety, equipment integrity, plant availability,
product quality, economics; an action trading a higher objective for a lower one
is inadmissible); diagnosis framed in terms of accumulation balances;
simulate-before-act; verify-before-shutdown (an excursion predicted to peak
below the limit is not grounds for shutdown); and restraint on self-correcting
disturbances. Part~B is a process description generated from the plant's
instrumentation and control configuration, its control-loop registry, validated
bounds, alarm table, and an input--output sensitivity table derived from the
nominal model; it contains the plant-specific content. Applying the layer to a
different process requires regenerating Part~B only. No task-specific training,
fine-tuning, or worked examples are used.

\textbf{Diagnosis and its scoring.} {\sloppy The ASM layer returns a diagnosis
from a closed twelve-item vocabulary: \texttt{feed\_composition\_upset},
\texttt{feed\_loss\_or\_header\_pressure}, \texttt{reactor\_cooling\_fault},
\texttt{condenser\_cooling\_fault}, \texttt{kinetics\_drift},
\texttt{heat\_transfer\_degradation}, \texttt{valve\_stiction\_or\_stuck},
\texttt{sensor\_fault}, \texttt{operating\_point\_change},
\texttt{compound\_fault}, \texttt{unknown}, and \texttt{none}.\par} Each scenario is
labelled in advance with one or two ground-truth tags from this vocabulary
(Supplementary Table~S4); \texttt{unknown} is scored incorrect on every
scenario, and \texttt{none} is the ground-truth tag only for Class~A. A
diagnosis is scored correct if it matches any
labelled tag, and the any-invocation diagnosis accuracy reported in
\cref{tab:headline} counts an episode correct if any of its invocations matches
a ground-truth tag; this is a best-of-$n$ metric. The rule is
fixed before the campaign and applied identically to the scripted supervisor.
Supplementary Table~S7 reports the single-prediction (final-diagnosis)
accuracy and confusion matrix for the model-supervised episodes; the
final-diagnosis accuracy is consistently lower than the any-invocation figure,
because the model commonly reports intermediate diagnoses before the correct
tag.

\textbf{Action interface.} Each invocation returns exactly one action: a
setpoint change to any of nine regulatory
setpoints within validated ranges, per-move limits and cooldowns; activation of
the composition trims or the override loops; one of four pre-audited procedures
(a 25\% or 50\% staged production cut, a depressurization step, or a
minimum-stable-rate park); no action; or an orderly shutdown, which counts as
failure on the primary endpoint. A programmatic validator applies
static checks (range, rate, cooldown, per-episode action budget) and, for
setpoint changes and reconfigurations, a shadow rollout on a dedicated
simulator instance over a 6-h horizon that vetoes actions predicted to trip the
plant or to breach margin floors; a rejected proposal returns machine-readable
reasons and the model may revise once.

\textbf{Forward simulation.} The model may call a what-if tool that
integrates the nominal closed loop forward from the current state under a
candidate action and returns whether the plant remains within constraints, the
time-to-trip, minimum margins, and product composition. The forward simulation
uses the nominal model with the current controller state. It is not given the
disturbance realization and is therefore an operator's simulator.

\subsection*{Models, experimental design, and statistics}

\textbf{Models.} Four configurations run the identical architecture, prompt,
process-state summary, action interface and validator: Claude Sonnet-5 (model identifier
\texttt{claude-sonnet-5}, Anthropic) with adaptive reasoning at high effort
(primary) and at low effort (the reasoning-effort comparison). GLM-5.2
(\texttt{z-ai/glm-5.2}) and DeepSeek-V4-Flash
(\texttt{deepseek/deepseek-v4-flash}) were accessed through the OpenRouter
inference service with the forward-simulation tool executed in-process (the
cross-model comparison). All runs were performed in July 2026. Reasoning effort
is defined operationally by the provider's thinking-token budget; the exact API
parameters for every configuration (sampling temperature, maximum output
tokens, thinking-token budget, and retry policy) are listed in Supplementary
Note~2. Model outputs
are stochastic; the campaign therefore runs three paired seeds per scenario,
reports every episode, and caches each response keyed by the canonical
request, so the reported campaign replays offline byte-identically. Independent re-sampling was
not performed. The
scripted supervisor is a deterministic rule table over the same process-state summary and
interface, written from generic alarm-response practice (shed production on
critical pressure, activate the standing protections on sustained alarms,
adjust the corresponding setpoint on level alarms) before the main campaign and
not tuned to individual scenarios; it receives no scenario labels. All decision
transcripts for every configuration are retained.

\textbf{Verification ablation.} Two configurations replicate the primary
configuration
(model, reasoning effort, prompt hash, process-state summary, action interface, scenarios and
paired seeds all identical) with the run-time verification progressively
removed: \emph{no-shadow} retains the static range, rate, cooldown and budget
checks but disables the forward-simulation veto; \emph{no-validator} disables
all checks, so every action the model proposes is applied as proposed. The
bounded action interface itself is unchanged (direct actuator manipulation
remains unrepresentable), and the model's optional what-if tool remains
available in both. Each spans 39 episodes, paired by (scenario, seed)
against the fully verified primary configuration.

\textbf{Endpoints.} Primary: the plant is stabilized within all hard
constraints at the horizon (interlock trip and orderly shutdown both
count as failure). Secondary: orderly shutdown versus trip, time-to-trip,
constraint-violation integrals, operating cost, off-specification production,
diagnosis accuracy against the ground-truth tag, accepted actions per episode,
decision latency, and model cost; the full set is reported in the Supplementary
Information without multiplicity adjustment, as these endpoints are descriptive.
A class-A non-inferiority check against the reference (no
degradation of the no-intervention class) was specified before the main
campaign; the pre-specified margin and its result are reported in the
Supplementary Information.

\textbf{Statistics.} The experimental unit is the episode $e = (\text{scenario},
\text{seed})$, and every contrast between configurations $A$ and $B$ is paired on $e$.
With discordant counts $b = \#\{e : S^A_e = 1, S^B_e = 0\}$ and
$c = \#\{e : S^A_e = 0, S^B_e = 1\}$, the exact (binomial) McNemar test reports
\begin{equation*}
p \;=\; \min\Bigl\{1,\; 2 \sum_{i=0}^{\min(b,c)} \binom{b+c}{i}
\Bigl(\tfrac{1}{2}\Bigr)^{b+c}\Bigr\},
\end{equation*}
and the paired difference in the in-constraint fraction is
$\hat{\Delta} = \tfrac{1}{n}\sum_e \bigl(S^A_e - S^B_e\bigr)$ with $n = 39$,
its 95\% confidence interval computed by Newcombe's method for paired
proportions (square-and-add of the two Wilson single-proportion limits with a
$\phi$-correlation correction from the paired $2\times2$ table). Continuous
endpoints use paired per-episode differences with cluster-bootstrap confidence
intervals: we resample scenarios with replacement, then seeds within
scenario, for 10{,}000 replicates, and report the percentile interval.
We report every episode: 156 model-supervised, 78 verification-ablation, and
117 non-model episodes (39 baseline, 39 reference controller, 39 scripted
supervisor). No episode was excluded.

\subsection*{Data and code availability}

The abnormal-situation management harness, the benchmark definition, the prompts (Parts A and B), the validator bounds, and the analysis workflow are publicly available in the project repository at \url{https://github.com/khalidlabs/safetep-core}. The Tennessee Eastman process kernel used is an independent implementation, publicly available at \url{https://github.com/khalidlabs/tep-studio}.

\section*{Funding}

No funding was received for this work.

\section*{Author contributions}

K.A. conceived and designed the study, developed the abnormal-situation
management architecture and the experimental harness, performed the
experiments and the analysis, prepared the figures, and wrote the manuscript.

\section*{Competing interests}

The author declares no competing interests.

\section*{Correspondence}

Correspondence and requests for materials should be addressed to K.A.
(hazmimm@sabic.com).

\bibliographystyle{unsrtnat}
\bibliography{refs}

\end{document}